\documentclass[twocolumns]{aa}

\usepackage{graphicx}
\usepackage{psfig}
\usepackage{epsfig}

\def\ltsima{$\; \buildrel < \over \sim \;$}
\def\simlt{\lower.5ex\hbox{\ltsima}}    
\def\gtsima{$\; \buildrel > \over \sim \;$}
\def\simgt{\lower.5ex\hbox{\gtsima}}    

\def\ref{\par\noindent\hangindent 20 pt}

\def\mincir{\ \raise -2.truept\hbox{\rlap{\hbox{$\sim$}}\raise5.truept 
\hbox{$<$}\ }}  %
\def\magcir{\ \raise -2.truept\hbox{\rlap{\hbox{$\sim$}}\raise5.truept %
\hbox{$>$}\ }}

\begin{document}

\title{VLT adaptive optics imaging of QSO  host galaxies and 
close environment at z $\sim$ 2.5: results from a pilot program. }

\author{Renato Falomo\inst{1}, J.K. Kotilainen\inst{2}, Riccardo Scarpa\inst{3} \and Aldo Treves\inst{4} }

\offprints{R. Falomo}

\institute{INAF -- Osservatorio Astronomico di Padova, Vicolo dell'Osservatorio 5, 
35122 Padova, Italy 
\email{falomo@pd.astro.it}
\and
Tuorla Observatory, University of Turku, V\"ais\"al\"antie 20, 
FIN--21500 Piikki\"o, Finland
\email{jarkot@utu.fi}
\and
European Southern Observatory, 3107 Alonso de Cordova, 
Santiago, Chile. 
\email{rscarpa@eso.org}
\and
Universit\`a dell'Insubria, via Valleggio 11, 22100 Como, Italy
\email{treves@mib.infn.it}
}

\date{Received; accepted }

\abstract{
We report on ESO--VLT  near-IR adaptive
optics imaging of one radio-loud (PKS 0113-283) and two radio quiet (Q
0045-3337 and Q0101-337) quasars at z $>$ 2.  
In the first case we are able to resolve the QSO and find that it is 
hosted by an elliptical of absolute magnitude M$_K = -27.6$. For 
the other two objects no extended emission has been unambiguously detected. 
This  result, though restricted  to a single object, 
extends up to z=2.5  the finding that cosmic
evolution of radio-loud quasar hosts follows the trend expected for
luminous and massive spheroids undergoing passive evolution.
For Q0045-3337 our high resolution images show that it is located
1.2 arcsec from  a K=17.5 foreground disc galaxy, which may act 
as gravitational lens, since the QSO most probably lies within the galaxy Einstein radius.

\keywords{Galaxies:active -- Infrared:galaxies -- Quasars:general --
galaxies: evolution }
}

\titlerunning{VLT imaging of z $\sim$ 2.5 QSO }
\authorrunning{ Falomo et al.}

\maketitle

\section{Introduction}

Ground­-based imaging (e.g., McLeod \& Rieke 1994; Taylor et al. 1996;
Kotilainen \& Falomo 2000; Percival et al. 2001) together with  higher
resolution HST data (e.g., Bahcall et al. 1997;
Hooper, Impey, \& Foltz 1997; Boyce et al. 1998; \cite{hutch99}; 
\cite{hamilton02}; Dunlop et al. 2003; Pagani,
Falomo, \& Treves 2003) clearly show that in the local universe
(z$<$0.5) powerful quasars  are found in massive galaxies
dominated by the spheroidal component.

At higher redshift (up to z $\sim$ 2)  the available data seem 
to confirm this scenario (e.g. \cite{hutch98}; \cite{hutch99}; Kukula et al. 2001; \cite{ridgway01}; 
\cite{falomo01}; \cite{falomo04}), in particular indicating that the luminosity of quasar host galaxies
is consistent with that of massive spheroids undergoing passive evolution.

Properties of quasar hosts at higher redshift are very poorly known because of
the severe difficulties of getting direct information on these galaxies 
(e.g. see the pioneering papers by 
\cite{hutch95}; \cite{lehnert92}; \cite{lowenthal}).
On the other hand  some indirect arguments supporting the above scenario for QSO host evolution 
at even higher redshift became  recently available.  The Sloan Digital Sky Survey 
has observed hundreds of high redshift quasars (Schneider
et al. 2003) sufficient to constraint their space  density up to $z\sim 6$ (e.g. Fan et
al. 2001, 2003).
 These objects trace the existence of $ \sim 10^9$ M$_\odot$  super massive BHs  (Fan et al. 2001,
2003; Willott, McLure \& Jarvis 2003). Therefore if the link between the BH mass and the hosting spheroid 
holds also at these epochs one could argue that massive host galaxies are already formed at these redshifts.
 Furthermore, evidence has been found also 
in the highest redshift objects for molecular gas (Bertoldi et
al. 2003b; Walter et al. 2004), dust (Bertoldi et al. 2003a) and
metals (Freudling, Corbin \& Korista 2003), indicating that
even at these early epochs quasars are likely associated with galaxies that 
have experienced significant star formation.

These observations  appear at odds with a cosmological $\Lambda$ cold dark matter  scenario, 
which  predicts a significant drop in mass of high z galaxies as a
consequence of the processes of hierarchical merging (e.g. \cite{kauffmann00};  
\cite{haehnelt04} ).

In this context it is therefore important to push as far as possible in redshift the direct
detection and characterization of QSO host galaxies. 
In particular, a key point is to probe the QSO host properties at
epochs close to (and possibly beyond) the peak of quasar activity (z
$\sim$ 2.5).  This is not attainable with HST because of the modest aperture that 
translates into a limited capability to detect faint extended nebulosity.  One has thus 
to resort to 10 meter class telescopes equipped with adaptive optics (AO) systems. This  keeps  
the advantage of both high spatial resolution and high
sensitivity although some complications are introduced (namely the need for a near reference point source and a 
PSF that can significantly vary spatially and temporally)
Moreover, unless artificial (laser) guide stars are available 
(not yet fully implemented in current AO systems) 
only targets which are sufficiently angularly close to relatively 
bright stars can actually be observed. 

Adaptive optics imaging of quasar hosts has been so far obtained for a small number of 
objects at relatively low redshift and using the first usable AO systems at 4m class telescopes 
(Hutchings et al 1998, 1999, \cite{marquez01}, \cite{lacy02} ). Only very recently AO imaging 
systems have become available at 8m class telescopes and can be used to image distant QSO 
with the full capability of spatial resolution and adequate deepness. 
A recent report by Croom et al (2004) presents a study of 9 z $\sim$ 2 quasars imaged 
with AO at Gemini North telescope,  
but they are able to resolve only one source at z=1.93. No quasars at z $>$ 2 have been till now 
clearly resolved using AO imaging systems.

In order to investigate  the properties of quasar hosts near the peak of QSO activity 
we have carried out a pilot
program to secure K band images of quasars in the redshift range 2 $<$
z $<$ 3 using the AO system at ESO VLT.  
Here we present the results for  one radio-loud (PKS 0113--283) and two radio-quiet (Q 0045-3337 and Q
0101-337) QSOs. These observations allow us to resolve the quasar host of a z $\sim$ 2.6 quasar and, 
albeit based on a single object, support the view that the cosmic evolution of QSO hosts, at least until z $\sim$ 2.5, 
follows that of inactive massive spheroids. 
Throughout this work we use H$_0$ = 70 km s$^{-1}$ Mpc$^{-1}$, $\Omega_m$ =
0.3, and $\Omega_\Lambda$ = 0.7. 

\begin{center}
\begin{figure*}  
\centerline{\hbox{\psfig{file=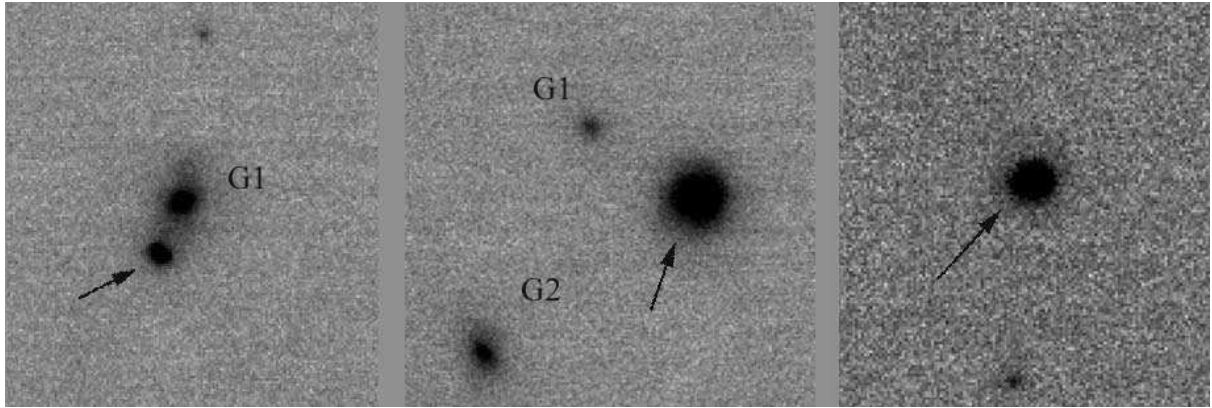,width=16cm} }}
\caption[vltnaco]{
The central portion (FoV: 6 $\times$ 6 arcsec; North is up and East to the left) of images  in the Ks filter 
of the three selected quasars 
(left: 0045-3337; middle:  0101-337; right: 0113-283).
At the redshift of the objects, these correspond to rest
frame $V$ or $R$-bands.  In all frames the target is indicated by the arrow. 
\label{vlt-naco1}}
\end{figure*}
\end{center}

\section{Object selection and observation strategy}

Adaptive optics systems employing natural guide stars imply that only
targets sufficiently  close to bright stars (used as
reference for AO correction) can be actually observed. 
Because of that, we searched the Veron \& 
Cetty-Veron (2001) catalog for quasars in the redshift range 2$<$ z $<$ 3 and
$\delta <$ 0, having a star brighter than V=14 within 30 arcsec.  In
these conditions the AO system at the VLT is expected to deliver
images of Strehl ratio better than $\sim$ 0.2 when the external seeing
is $<$0.6 arcsec.  The twenty candidates fulfilling these 
requirements were then inspected individually looking at 
Digitized Sky Surveys red plates obtained with the 
UK Schmidt Telescope and available at ESO. 
A priority was assigned on the basis of the quality of the guiding
star (magnitude and distance from the target).  For the given
allocated time we have then chosen one radio loud and 2 radio quiet
objects (Table 1).

When using AO systems the observed field of view is usually not large
enough to include stars that can be used to directly
evaluate the PSF, as it is often done in standard imaging.  Moreover, the
shape of the PSF may vary with  time (seeing and performances of
the system may change) and with the angular distance from the AO guide
star.  A possibility to characterize the PSF is to use the
information from the guide star.  This however can provide only a
limit to the PSF shape as the actual PSF at the position of the target
will be degraded.  If the guide star is very bright as required by the AO
technique to obtain optimal AO correction, saturation problems can be
encountered.  In order to cope with these problems we selected a number of 
pairs of stars with
angular distance similar to that between the targets and the star used
for AO correction that should be observed just before and/or after the
observations of the quasars.  These data would also provide 
a frame of reference to evaluate the spatial and temporal variations of the 
PSF. A similar procedure was adopted by \cite{croom04} for AO imaging of quasars at 
the Gemini telescope.

Unfortunately, this part of the program was not 
executed hampering the fulfillment of our analysis procedure. Thus it
was only possible to fully characterize the PSF for 
Q0113-283, which happens to have suitable stellar objects in the
frame.  For the other two cases no unambiguous method to evaluate the
PSF a posteriori was available so we have limited our analysis of these objects to the
environments.

With a significantly larger sample of QSO now available (as those produced by the 
Sloan Digital Sky Survey (SDSS) and 2dF surveys; 
\cite{schneider03} \cite{croom01}) one can select targets having secondary 
PSF stars in the observed field such 
as to allow a characterization  of the PSF using the same frame.

\section{Observations and data analysis}

The observations were obtained using NAOS (Rousset et al. 2002), the
first Adaptive Optics System on the VLT at the European Southern
Observatory (ESO) in Paranal (Chile). NAOS feeds CONICA, a 1 to 5 $\mu$m
imaging, chronographic, spectroscopic, and polarimetric instrument
(Lenzen et al. 1998).   
The K-band images were secured in service mode by ESO staff and are
detailed in Table 1. A dichroic sent the visible light to the visible
wave-front of NAOS, and the IR light to CONICA.  The CONICA detector
is an Aladdin InSb 1024 x1024 pixel array.  
The S54 camera provides a field of view of 56x56 arcsec with a sampling of 54
mas/pixel. 
The objects were observed using a jitter procedure and individual
exposures of 2 minutes per frame, for a total integration time of 38
minutes.  

Data reduction was performed by the ESO pipeline for jitter imaging
data (\cite{devillard99}). The normalized flat field was obtained by
subtracting ON and OFF images of the illuminated dome, after
interpolating over bad pixels. Sky subtraction was obtained by median
averaging sky frames from the nearest in time frames .  The reduced
frames were then aligned to sub-pixel accuracy and co-added.  
Photometric calibration, performed using standard stars
observed during the same night, yields an internal 
accuracy of  $\pm$0.1 mag.

%
\begin{table*}
\caption[]{Journal of the observations}
\begin{tabular}{lcccclccccc}

{Quasar} & {Type} & {z} &  {Date} & {V$^{a}$ } &
{$ t_{exp}^{b}$}  & {Seeing$^{c}$}  & {FWHM$^{d}$} & {K$^{e}$} & {GS(V)$^{f}$} & {GS(d)$^{f}$}\\
\hline
~Q 0045--333    & RQQ &2.140   & 17/Oct/2002  & 17.60 & 38 & 0.5 & 0.13 & 18.75 & 12.7 & 27 \\
~Q 0101--337    & RQQ &2.210   & 17/Oct/2002  & 15.95 & 38 & 0.5 & 0.26 & 18.10 & 10.1 & 28 \\
~PKS 0113--283  & RLQ &2.555   & 21/Oct/2002  & 17.10 & 38 & 0.6 & 0.22 & 20.02 & 13.5 & 15 \\
\hline
\end{tabular}
\\
$^{a}$  V magnitude from the \cite{veron01} catalogue. \\
$^b$ Total exposure time in minutes. \\
$^c$ External seeing in the optical band (arcsec). \\
$^d$ Image quality measured as the full-width at half maximum in the K-band (arcsec).\\
$^e$ Observed  K magnitude of the target. \\
$^f$ Guide star V magnitude and distance in arcsec from the target. 
\end{table*}

\section{Results for individual objects}

{\bf Q 0045-3337} is a radio quiet-quasar discovered by a prism
objective survey (\cite{iovino96}).  Our image (Fig. 1) shows the
presence of a companion galaxy (G1) very close (1.2 arcsec) to the
QSO. The rest of the observed field is rather empty.  The QSO image is
slightly elongated (ellipticity = 0.08) at PA = 230 degrees
(Fig. 2). Because of  the lack of other stellar objects in the field 
it is not possible to asses if this is a real effect. We note that while 
no elongation is seen in
the image of the guide star (GS) the direction of the elongation is near to the radius 
vector to the GS where possible asymmetries are expected by AO imaging.  
Since the radial brightness profile of the QSO is not more extended than that of the 
PSF of Q 0113-337 (see
below) we  assume that Q 0045-33 is unresolved in our image.

\begin{center}
\begin{figure}
\centerline{\hbox{\psfig{file=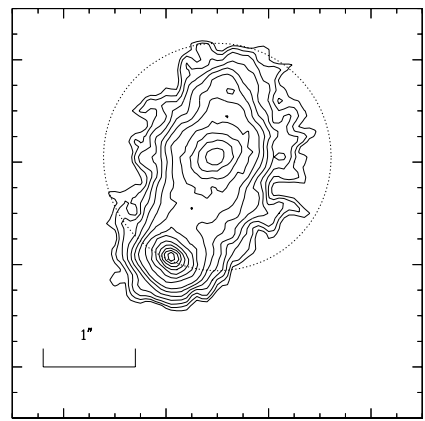,width=6.5cm} }}
\caption[q0045c]{Contour plot of the K band image of Q 0045-3337
(North is up and East to the left).  The QSO is the bright compact
source 1.2 arcsec South of the more extended,  K= 17.6, foreground galaxy.
The image of the quasar is slightly elongated along the
direction perpendicular to the QSO-galaxy pair.  The  circle 
represents the Einstein ring centered on the companion galaxy for a
point source of mass M = 1  $\times$ 10$^{11}$ M$_\odot$ (see
text).
\label{q0045c}}
\end{figure}
\end{center}

The companion galaxy is  well represented by a disk model
with a half light radius of 0.4 arcsec and a total magnitude K =
17.6.  For a reasonable range for its absolute magnitude (M$_K$ =
-24 to -26) and the effective radius (2-3 kpc)  we
estimate that the redshift is between 0.4 and 1.  For a
conservative value of the mass -to-light ratio M/L $\sim$ 3 this implies that the companion
galaxy has a mass M $\approx$ 1 to 5 $\times$ 10$^{11}$ M$_\odot$.  At
these redshifts such a galaxy would have an Einstein radius 
$\alpha_E$ in the range 1.2
to 2.6 arcsec, comparable or larger than the observed angular
separation between the QSO and the foreground galaxy (see Fig. 2).  
Under these circumstances it is possible that 
the galaxy acts as a gravitational lens on the QSO, and one would expect
to see multiple images of the quasar (e.g. Jetzer 2002). Apart from the
possibility that the object at $\sim$ 4.5 arcsec N is due to
a very demagnified secondary image of the QSO we have no indications from our frame of
other secondary images.
 
A further study of this system obviously requires the determination of
the redshift of the lens (and thus a better estimate of the mass) in order to asses the 
lensing effects of the foreground galaxy.

\begin{center}
\begin{figure}
\centerline{\hbox{\psfig{file=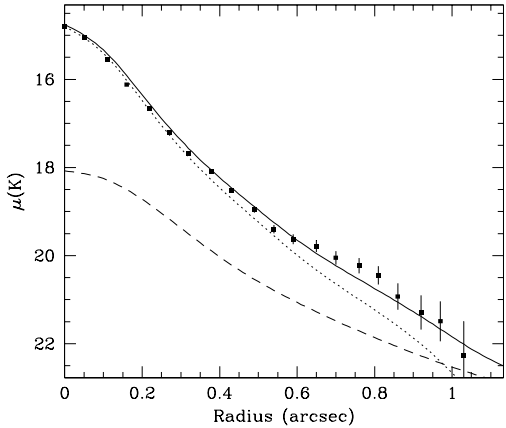,width=7cm} }}
\caption[q0113p]{The observed radial brightness profiles of the QSO 0113-283  
(filled squares), superimposed to the fitted model (solid line) consisting of the PSF 
(dotted line) and an elliptical  galaxy convolved with 
its PSF (dashed line). The associated errors are a combination of the statistical 
photometry in each bin and of the uncertainty on the background level 
(that is dominant at the faintest fluxes). 
\label{q0113p}}
\end{figure}
\end{center}

{\bf Q 0101-337} was classified  as a radio quiet quasar discovered by visual 
search for emission lines and
UV continuum objects on UKST objective prism plates
(\cite{savage84}).  The source is not detected in the 2MASS
(Barkhouse \& Hall 2001) but it appears in the NRAO VLA Sky Survey 
at a flux level of 27.1 mJy at 1.4 GHz.

The light profile of this QSO exhibits a substantial tail at radii
larger than 0.5 arcsec with shape significantly different from that of
point like sources in the other images.  This may indicate either that the
object is resolved, or that the AO correction was not optimal and produced 
a PSF with an extended wing.
Unfortunately, the lack of stars in the field of view prevent us from
accurately model the PSF, thus it was not possible to discriminate
between these two possibilities. Conservatively, we  assume also in this case the
target is unresolved.  The QSO is surrounded by a number of faint
galaxies. In Fig. 1 we show the two closest ones: G1 with K= 19.0 at 3
arcsec NE (projected distance 25 kpc) and G2 with K= 17.6 at 6.3 arcsec SE.

\begin{center}
\begin{figure} 
\centerline{\hbox{\psfig{file=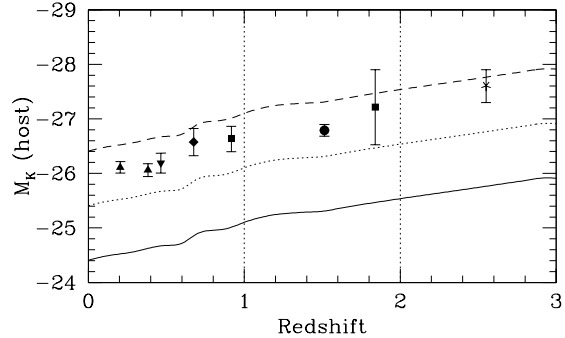,width=8.5cm} }}
\caption[rlqevol]{ The evolution of radio loud quasar host luminosity
compared with that expected for massive ellipticals (at M$^*$, M$^*$-1
and M$^*$-2; {\it solid, dotted and dashed} line ) undergoing passive
stellar evolution (\cite{bressan98}).  The new detected host galaxy
at z $\sim$ 2.5 (asterisk) is compared with the data for samples of
lower redshift RLQ presented in Falomo et al 2004.  Each point is
plotted at the mean redshift of the considered sample while the error
bar represents the $1\sigma$ dispersion of the mean except for the new detection 
where the  uncertainty of the measurement is given.
\label{rlqevol}}
\end{figure}
\end{center}

{\bf PKS 0113-283} 
is a flat spectrum radio loud quasar at z=2.555
(Hook et al 2003 ).  In the observed field there is a secondary star
(K=15.7) at approximately the same angular distance (16.3 arcsec) from the AO guide
star as the target (which is at 15 arcsec from the GS). 
This is an ideal situation for deriving a reliable
PSF since the shape of the PSF mainly depends on the distance from the AO guide star. 
Possible 2D asymmetries of the PSF due to difference 
in the relative positions of the target and the PSF star with respect to the the AO guide star 
have negligible effect when one considers the azimuthal averaged radial profile. 
The comparison between the radial brightness profile of this
star with that of the QSO indicates that the target is clearly
resolved (see Fig. 3). A detailed modeling of the luminosity profile using an
iterative least-squares fit was then performed, assuming a combination
of a point source and an elliptical galaxy convolved with the proper
PSF.  The radial brightness profile of the quasar compared with that 
of the PSF star is shown in figure 3. 
The result of the best fit (Figure 3) indicates that the host galaxy is a 
luminous giant elliptical: K=19.1 $\pm$ 0.3  corresponding to M$_K$ = -27.6 
(including K-correction) and an effective radius Re = 7.5 $\pm$ 3 kpc.  
These properties are in the range of the observed values for
 inactive ellipticals in clusters and groups (e.g. \cite{pahre98} ). 
According to the relation between K-band luminosity and R$_e$ define by these authors 
 for R$_e \sim$ 7 kpc one expects M$_K$ is  between -25 and -28.5  in agreement with
 our measurement of the quasar host (M$_K$ = --27.6 or M$_K \sim$ --26.5 assuming 1 mag 
evolution effect).

\section{Conclusions}

We have presented the results of a pilot project aimed at obtaining
adaptive optics near-IR images of quasars at z $\sim$ 2 -- 3. We showed  that 
when a PSF characterization is available,  
it is possible to directly detect the host galaxy up to these high
redshifts.  
 The  detection of a quasar host at z $\sim$ 2.6 confirms
the trend (e.g. \cite{falomo04} ) of the host luminosity  of RLQ 
observed for lower redshift objects. This new measurement 
 taken, together 
with the available data on quasar hosts for  lower redshift objects (see Fig 4), 
indicate that up to quite high redshift, radio loud quasars are  hosted by 
massive fully formed galaxies. 
This  result is consistent with the observed trend of luminosity of 
radio galaxies (\cite{willott03} )
and suggests that up to z=2.5  there is no evidence 
of the decrease in mass (luminosity) as expected in the hierarchical merging 
scenario for the formation and evolution of spheroidal galaxies. 
The models for the joint formation and evolution of massive galaxies and their supermassive 
black holes  (e.g. \cite{kauffmann00} ) in fact  predict that at high redshift 
quasars should be found in progressively less massive galaxies. 

Finally we note that the serendipitous discovery of a galaxy at $\sim$ 1 arcsec
from one of our quasars has yielded an unexpected gravitational lens
candidate, showing the potentialities of AO in this context as well.
The disc morphology of the candidate lens makes this system of particular interest and 
requires further investigation since it may belong to 
a tiny minority of disk galaxy lenses (\cite{winn03}).  

\section*{Acknowledgments}
We are grateful to Matteo Chieregato for helpful discussions on the possible lensing effects on the 
object Q0045-3337. 
This work was partially supported by the ASI-IR 073 and ASI-IR 056, 
 FIN-INAF  2003 and by the Academy of Finland (project 8201017). 
This research has made use of the NASA/IPAC Extragalactic Database {\em(NED)} 
which is operated by the Jet Propulsion Laboratory, 
California Institute of Technology, under contract with the 
National Aeronautics and Space Administration.


\begin{thebibliography}{}

\bibitem [Bahcall et al. 1997]{B97} Bahcall, J.N., Kirhakos S., Saxe D.H., Schneider D.P. 1997, ApJ 479, 642
\bibitem [Barkhouse et al 2001] {barkhouse01}Barkhouse W.A. and Hall P.B., 2001, AJ, 121, 2842
\bibitem [Bertoldi et al. 2003a]{bertoldi03a} Bertoldi F., Carilli C.L., Cox P.,
	et al. 2003b A\&A 406, L55
\bibitem [Bertoldi et al. 2003b]{bertoldi03b} Bertoldi F. et al. 2003a 
	A\&A 409, L47
\bibitem [Boyle et al. 2000]{boyle00} Boyle, B.J., Shanks, T., Croom, S.M., Smith, R.J., Miller, L., Loaring, N., Heymans, C. 2000,
			MNRAS, 317, 1014
\bibitem  [Bressan Granato \& Silva 1998]{bressan98} Bressan, A., Granato G.L. \& Silva L. 1998, A\&A 332, 135.
\bibitem [Croom et al 2001] {croom01} Croom S.M., Smith R.J., Boyle B.J., Shanks T., Loaring N.S., Miller L.,
    Lewis I.J. 2001 MNRAS 322, L29 
\bibitem [Croom et al 2004] {croom04} Croom S.M., Schade D., Boyle B.J., Shanks T., Miller L., Smith R.J. 
2004 ApJ 606 126
\bibitem [Devillard 1999]{devillard99} Devillard, N., 1999, Astronomical Data Analysis Software and Systems VIII, ASP Conference Series, Vol. 172 (ed. D.M. Mehringer, R.L. Plante, D.A. Roberts), p. 333
\bibitem [Dunlop \&  Peacock 1990]{dunlop90}  Dunlop, J. S., Peacock, J. A. 1990 MNRAS, 247,19
\bibitem [Falomo et al 2001]{falomo01} Falomo R., Kotilainen J.K., Treves A. 2001, ApJ 547, 124
\bibitem [Falomo et al 2004]{falomo04} Falomo, R. Kotilainen, J.K., Pagani, C. Scarpa R. \& Treves A. 2004, ApJ 604 495
\bibitem [Fan et al. 2001]{fun01} Fan X. et al. 2001, AJ, 121, 54
\bibitem [Fan et al. 2003]{fun03} Fan X. et al. 2003, AJ, 125, 1649
\bibitem [Freudling, Corbin \& Korista 2003]{freudling03} Freudling W., 
	Corbin M.R., \& Korista K.T. 2003 ApJ 587, 67
\bibitem [Haehnelt 2004]{haehnelt04}  Haehnelt, M., 2004, 
Carnegie Observatories Astrophysics Series, Vol. 1: Coevolution of Black Holes and Galaxies, 
ed. L. C. Ho (Pasadena: Carnegie Observatories, 
http://www.ociw.edu/ociw/symposia /series/symposium1/proceedings.html)
\bibitem [Hamilton, Casertano \&Turnshek 2002] {hamilton02} Hamilton T.S. Casertano, Turnshek D.A. 2002 ApJ 576, 61
\bibitem [Hook et al ] {hook03}  Hook M.,   Shaver P. A.,  Jackson C. A.,  Wall J. V., Kellermann  K. I. 2003 
A\&A 399 469
\bibitem [Hooper et al. 1997]{hooper97} Hooper, E.J., Impey C.D. \& Foltz C.B., 1997, ApJ, 480, L95
\bibitem [Hutchings 1995]{hutch95} Hutchings, J. B. 1995 AJ, 110, 994
\bibitem [Hutchings 1998]{hutch98} Hutchings, J. B. 1998 AJ, 116, 20
\bibitem [Hutchings et al 1998]{hutchings98} Hutchings J.B. Crampton D., Morris S.L., Steinbring E. 1998 
PASP 110 374
\bibitem [Hutchings et al 1999]{hutch99} Hutchings J.B., Crampton D., Morris S.L.,Durand D., Steinbring E., 1999
                AJ 117, 1109
\bibitem [Iovino Clowes  \& Shaver 1996]{iovino96} Iovino A.,Clowes R.,Shaver P. 1996,A\&AS 119,265 
\bibitem [Jetzer 2002]{jetzer02} Jetzer P. 2002 in Modern Cosmology, eds. Bonometto S. Gorini V. \& Moschella U., 
IOP, Bristol \& Philadelfia p.378.
\bibitem [Kauffmann \& Haehnelt 2000]{kauffmann00} Kauffmann, G., Haehnelt, M., 2000, MNRAS, 311, 576

\bibitem [Kotilainen \& Falomo 2000]{kotifal00} Kotilainen, J.K., Falomo, R. 2000,  A\&A, 364, 70
\bibitem [Kukula et al. 2001] {kukula01} Kukula M.J., Dunlop J.S.,McLure R.J., Miller L., Percival W.J., Baum S.A
                O'Dea, 2001, MNRAS, 326, 1533
\bibitem [Lacy et al 2002] {lacy02} Lacy M., Gates, E.L. Ridgway, S.E., de Vries W. Canalizo G., Lloyd, J.P., 
	Graham, J. R. 2002 AJ 124 3023
\bibitem [Lehnert et al. 1992] {lehnert92} Lehnert, M.D., Heckman, T.M., Chambers, K.C., Miley, G.K. 1992 ApJ, 393, 68
\bibitem [Lenzen et al. 1998] {lenzen98}  Lenzen R., Hofmann R., Bizenberger P., Tusche, A., 1998, Proc. SPIE, 3354,
606
\bibitem [Lowenthal et al. 1995]{lowenthal} Lowenthal, J.D., Heckman, T.M., Lehnert, M.D., Elias, J.H. 1995 ApJ, 439, 588
\bibitem [Marquez et al 2001] {marquez01} Marquez I., Petitjean P., Théodore B., Bremer M., Monnet G., Beuzit J.-L. 2001
AA 371 97
\bibitem [McLeod \& Rieke 1994]{mcleod94} McLeod, K.K., Rieke, G.H., 1994, ApJ, 431, 137
\bibitem [Pahre et al. 1998]{pahre98} Pahre, M. A. Djorgovski, S. G. de Carvalho, R. R. 1998 ApJ 116 1591
\bibitem [Pagani et al. 2003]{pagani03} Pagani, C., Falomo, R., Treves, A., 2003, ApJ 596, 830
\bibitem [Percival et al. 2000]{perci00} Percival, W.J., Miller, L., McLure, R.J., Dunlop, J.S. 2000, MNRAS 322, 843
\bibitem [Ridgway et al. 2001]{ridgway01} Ridgway, S., Heckman, T., Calzetti, D., Lehnert, M. 2001, ApJ, 550, 122
\bibitem [Rousset et al. 2002] {rousset02} Rousset, G., Lacombe, F., Puget, P., Gendron, E., Arsenault, R., et al., 2000,
Proc. SPIE, 4007, 72
\bibitem [Savage et al 1984] {savage84} Savage A., Trew A.S., Chen J., Weston T. 1984 MNRAS. 207 393
\bibitem [Schneider et al. 2003]{schneider03} Schneider D.P.  Fan X., Hall P.B., Jester S., Richards G.T., Stoughton C.,
    Strauss M.A., Subbarao M., Vanden Berk D.E., Anderson S.F., et al. 2003, AJ, 126, 2579
\bibitem [Taylor et al. 1996]{taylor96} Taylor, G.L., Dunlop, J.S., Hughes, D.H., Robson, E.I. 1996, MNRAS, 283, 930
\bibitem [Veron-Cetty \& Veron 2001] {veron01} Veron-Cetty M.P., Veron P. 2001, A\&A, 374, 92
\bibitem [Walter et al. 2004]{walter04} Walter F. et al. 2004, Nat, in press 	(astro-ph 0307410)
\bibitem [Willott, McLure \& Jarvis 2003]{willott03} Willott C.J.,  McLure R.J. 
	\& Jarvis M.J. 2003 ApJ 587, L15
\bibitem [Winn et al 2003] {winn03} Winn, J.N., et al. 2003, ApJ 697, 672.
\end{thebibliography}
\end{document}